\documentclass[epj]{svjour}
\usepackage{graphics}
\usepackage{bm}
\newcommand{\G}{{\cal G}}
\newcommand{\x}{{\bm x}}

\newcommand{\be}{\begin{equation}}
\newcommand{\ee}{\end{equation}}
\begin{document}
\title{Capillary filling using Lattice Boltzmann Equations: the case of multi-phase flows}
\author{ F. Diotallevi \inst{1}  \and L. Biferale\inst{2} \and S. Chibbaro \inst{3} \and  A. Lamura \inst{4} \and G. Pontrelli \inst{1} \and  M. Sbragaglia\inst{5} \and S. Succi  \inst{1} \and F. Toschi \inst{1}
}                     
\offprints{biferale@roma2.infn.it}          
\institute{Istituto per le Applicazioni del
  Calcolo CNR, Viale del Policlinico 137, 00161 Roma, Italy. \and Dept. of Physics and INFN, University of Tor Vergata, Via della Ricerca Scientifica 1, 00133 Roma, Italy. \and Dept. of Mechanical Engineering, University of "Tor Vergata", Viale
Politecnico 8, Rome, Italy. \and Istituto per le Applicazioni del
  Calcolo CNR, Via Amendola 122/d 70126 Bari, Italy. \and Department of Applied Physics, University of Twente, P.O. Box 217, 7500 AE Enschede, The Netherlands.}
\date{Received: date / Revised version: date}
%
\abstract{ We present a systematic study of capillary filling for
  multi-phase flows by using mesoscopic lattice Boltzmann models describing a diffusive
  interface moving at a given contact angle with
  respect to the walls. We compare the numerical results at changing
  the density ratio between liquid and gas phases, $\delta \rho/\rho$
  and the ratio, $\delta \xi / H$, between the typical size of
  the capillary, $H$, and the interface width, $\delta \xi$. 
  It is shown that numerical results yield quantitative agreement with the
  Washburn law when both ratios are large, i./e. as the
  hydrodynamic limit of a infinitely thin interface is approached. 
  We also show that in the initial stage of the filling process, transient behaviour 
  induced by inertial effects and ``vena contracta'' mechanisms, may induce 
  significant departure from the Washburn law. 
  Both effects are under control in our lattice Boltzmann
  equation and in good agreement with the phenomenology of capillary filling.
} 
\maketitle

\section{Introduction}
\label{intro}
The physics of capillary filling is an old problem, originating with
the pioneering works of Washburn \cite{washburn} and Lucas
\cite{Lucas}. Recently, with the explosion of theoretical,
experimental and numerical works on microphysics and nanophysics, the
problem attracted a renewed interest
\cite{degennes,dussain,washburn_rec,tas}.  Capillary filling is a
typical ``contact line'' problem, where the subtle 
non-hydrodynamic effects taking place at the contact point between liquid-gas
and solid phase allows the interface to move, pulled by capillary
forces and contrasted by viscous forces. Usually, only the late
asymptotic stage is studied, leading to the well-known Washburn 
law, which predicts the following relation for the position of the 
interface inside the capillary:
\begin{equation}
\tilde z^2(t) - \tilde z^2(0) = \frac{\gamma H cos(\theta)}{3 \mu } \tilde  t 
\end{equation}
where $\gamma$ is the surface tension between liquid and gas, $\theta$
is the {\it static} contact angle, $\mu$ is the liquid viscosity, $H$
is the channel height and the factor $3$ depends on the geometry of
the channel (here a two dimensional geometry given by two infinite
parallel plates separated by a distance $H$ -- see fig. \ref{fig:1}). 
The above expression can be recasted in dimensionless 
variables $t=\tilde t/t_{cap}$ and $z = \tilde z /H$, being the
capillary time $t_{cap} = H \mu/\gamma$.
This leads to the universal law:
\begin{equation}
z^2(t) - z^2(0) = \frac{cos(\theta)}{3} t.
\label{eq:washburn}
\end{equation}
As already remarked in many works \cite{washburn_rec}, the asymptotic
behaviour (\ref{eq:washburn}) is obtained under the assumption that
(i) the inertial terms in the Navier-Stokes evolution are negligible,
(ii) the instantaneous {\it bulk} profile is given by the Poiseuille
flow, (iii) the microscopic slip mechanism which allow for the
movement of the interface is not relevant for bulk quantities (as the
overall position of the interface inside the channel), (iv) inlet and
outlet phenomena can be neglected (limit of infinitely long channels);
(v) the liquid is filling in a capillary, either empty or filled with
gas whose total mass is negligible with respect to the
liquid one. In the following, we will address all these effects and 
show to which extent they can described by using a mesoscopic
model for multiphase flows based on the discretized version of
Boltzmann Equations in a lattice. The model here used is a suitable
adaptation of the Shan-Chen pseudo-potential LBE \cite{shanchen} with
hydrophobic/hydrophilic boundaries conditions, as developed in
\cite{prenoi1,prenoi2}. Other models with different boundary conditions and/or
non-ideal interactions have been also used in \cite{kopo}.

\section{LBE for capillary filling}
\begin{figure}
  \resizebox{0.45\textwidth}{!}{
    \includegraphics{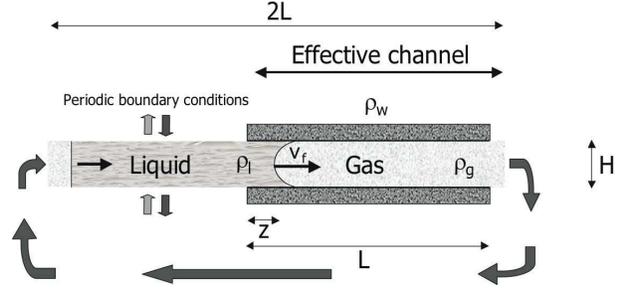}
  }
   \caption{Geometrical set-up of the numerical LBE. The 2 dimensional
     geometry, with length $2L$ and width $H$, is divided in two
     parts. The left part has top and bottom periodic boundary
     conditions such as to support a perfectly flat gas-liquid
     interface, mimicking a ``infinite reservoir''. In the right
     half, of length $L$, there is the true capillary:  the top and
     bottom boundary conditions are those of a solid wall, with a
     given contact angle $\theta$ \cite{prenoi1}. Periodic boundary conditions are also imposed at the west and east sides. } 
   \label{fig:1}
\end{figure}

The geometry we are going to investigate is depicted in
fig. (\ref{fig:1}). The bottom and top surface is coated only in the
rigth half of the channel with a boundary condition imposing a given
static contact angle \cite{prenoi1}; in the left half we impose
periodic boundary conditions at top and bottom surface in order to
have a flat liquid-gas interface which should mimic an ``infinite
reservoir''. Periodic boundary conditions are also imposed at the two
lateral sides such as to ensure total conservation of mass inside the
system. 

\subsection{LBE algorithm for multi-phase flows}
We start from the usual lattice Boltzmann equation with a single-time
relaxation \cite{Gladrow,Saurobook}: \be\label{eq:LB}
f_{l}(\bm{x}+\bm{c}_{l}\Delta t,t+\Delta
t)-f_l(\bm{x},t)=-\frac{\Delta t}{\tau_B}\left(
  f_{l}(\bm{x},t)-f_{l}^{(eq)}(\rho,\rho {\bm u}) \right) \ee
where $f_l(\bm{x},t)$ is the kinetic probability density function
associated with a mesoscopic velocity $\bm{c}_{l}$, $\tau_B$ is a mean
collision time (with $\Delta t$ a time lapse), $f^{(eq)}_{l}(\rho,\rho
{\bm u})$ the equilibrium distribution, corresponding to the
Maxwellian distribution in the continuum limit.  From the kinetic
distributions we can define macroscopic density and momentum fields as
\cite{Gladrow,Saurobook}: 
\begin{equation} \rho(\x)=\sum_{l} f_{l}(\x); \qquad  \rho
{\bm u}(\x)=\sum_{l}{\bm c}_{l}f_{l}(\x).\end{equation}   For technical details
and numerical simulations we shall refer to the nine-speed,
two-dimensional $2DQ9$ model \cite{Gladrow}. The equilibrium distribution in
the lattice Boltzmann equations is obtained via a low Mach number
expansion of the equilibrium Maxwellian \cite{Gladrow,Saurobook}.  In
order to study non-ideal effects we need to supplement the previous
description with an interparticle forcing. This is done by adding  a
suitable $F_l$ in (\ref{eq:LB}).  In the original  model
\cite{shanchen}, the bulk interparticle interaction is proportional to
a free parameter (the ratio of potential to thermal energy), ${\cal
  G}_{b}$, entering the equation for the momentum balance:
\be\label{forcing} F_{i}=-{\cal G}_{b} c^{2}_{s}\sum_l w(|{\bm
  c}_{l}|^2) \psi(\x,t) \psi (\x+{\bm c}_l\Delta t,t) {c}^{i}_l \ee
being $w(|{\bm c}_{l}|^2)$ the static weights
for the standard case of 2DQ9 \cite{Gladrow} and
$\psi(\x,t)=\psi(\rho(\x,t))$ the pseudo-potential function which
describes the fluid-fluid interactions triggered by inhomogeneities of
the density profile (see \cite{shanchen,prenoi1,prenoi2} for details).

One may show \cite{prenoi1,prenoi2} that the above pseudo-potential,
leads to a non-ideal pressure tensor given by (upon Taylor expanding the
forcing term): \begin{eqnarray} 
  \label{TENSORE}& P_{ij}&=\left(
    c^{2}_{s}\rho+{\G}_{b}\frac{c_{s}^{2}}{2}\psi^{2}+
    {\G}_{b}\frac{c^{4}_{s}}{4}|{\bf
      \nabla}\psi|^{2} +{\G}_{b} \frac{c^{4}_{s}}{2}\psi \Delta \psi
  \right) \delta_{ij}-  \nonumber \\ &-& \frac{1}{2}{{\G}_{b} {{c}}^{4}_{s}}\partial_{i}
  \psi \partial_{j}\psi+ {\cal O} (\partial^4), \end{eqnarray} where
$c_s$ is the sound speed. 
This approach allows the definition of a static contact angle $\theta$, by introducing  at the walls a
suitable value for the pseudo-potential $\psi(\rho_w)$ \cite{prenoi1},
which can span the range $\theta \in [0^o:180^o]$.
Moreover, it also defines a specific value for the surface tension, 
$\gamma_{lg}$, via the usual integration of the offset between normal and transverse
components of the pressure tensor along the liquid-gas interface
allows for\cite{shanchen,prenoi1,prenoi2}.\\
As to the  boundary conditions on the Boltzmann populations,
the standard bounce-back rule is imposed. One can show that the 
bounce-back rule gives no-slip boundary conditions up to second order
in the Knudsen number 
in the hydrodynamical  limit of single phase flows \cite{noijfm}. 
In presence of strong density variation, close to the walls across the 
interface, the velocity parallel to the wall may develop a small slip 
length (of the order of the interface thickness, $\lambda_s \propto
\delta \xi$) which in turn, allows for the interface to move. It is difficult to control exactly the phenomenon,
because even imposing an exact no-slip  boundary conditions at the wall \cite{yeomans}, the model will develop non 
trivial dynamics at the first node away from the wall, where both condensation/evaporation phenomena and/or spurious currents may conspire, leading to 
an overall non-zero slip velocity. For the scope of controlling the capillary 
filling, one may reabsorb all these effects within the usual Maxwell slip 
boundary conditions: $ u_{s} = \lambda_s \partial_{n} u$. 
It is easy to show that in presence of a slip velocity, the Poiseuille profile becomes:
\begin{equation}
\label{eq:profile}
u(y) = 6 \frac{ \bar{u} }{H^2} \frac{y(H-y)+\lambda H}{1 + 6 \lambda_s/ H}
\end{equation}
where the velocity of the front must be identified with the mean
velocity, $\bar{u} = 1/H \int_0^H dy u(y) = \dot z$.  Therefore, Washburn law
(\ref{eq:washburn}) becomes:
\begin{equation}
z(t)^2 - z_0^2 = \frac{cos(\theta)}{3}(1 + 6 \frac{\lambda_s}{H})  t 
\label{eq:washburn1}
\end{equation}

\subsection{Corrections to the Wahsburn law}
As already remarked many yeas ago \cite{bousanquet}, the Wahsburn law
(\ref{eq:washburn}) can be valid only if inertial forces are
negligible with respect to the viscous and capillary ones.  This
cannot be true at the beginning of the filling process, where strong
acceleration drives the interface inside the capillary.  However,
putting reasonable numbers for a typical microdevices experiments with
water ($H
\simeq 1 \mu m$, $\gamma \simeq 0.072 N/m$, $\rho_l \simeq  10^{-3} kg/m^3$, $\mu
\simeq 10^{-3} Ns/m^2$), one realizes that the transient time, $\tau_{diff} =
H \gamma \rho_l/\mu^2$, is usually very small, of the order of a few
nanoseconds, and therefore negligible for most experimental purposes.
Another important effect which must be kept in mind when trying to simulate
capillary filling, is the unavoidable ``resistance'' of the gas
occupying the capillary during the liquid invasion. 
This is a particular ``sensitive'' question, because reaching the typical $1:1000$ density ratio between
liquid and gas of experimental set up represents a challenge for most numerical methods, particularly
for multiphase Lattice Boltzmann which typically operates 
with density ratios of the order $1:10$ or $1:100$. 
In order to take in to account both effects, inertia and gas dynamics,
one may write down the balance between the total momentum change
inside the capillary and the force acting on the liquid$+$gas system:
\begin{equation}
\frac{d (\dot z M(t))}{dt} = F_{cap}+F_{vis}
\label{eq:momentum}
\end{equation}
where  $M(t)=M_g+M_l$ is
the total mass of liquid and gas inside the capillary at any given
time. The two forces in the right hand side correspond to the capillary force,
$F_{cap} = 2 \gamma cos(\theta)$, and to the viscous drag $F_{vis} =
-2 (\mu_g(L-z) + \mu_lz)\partial_nu(0)$. Following the notation of
fig.(\ref{fig:1}) and the expression for the velocity profile
(\ref{eq:profile}) one obtains the final expression (see also
\cite{napoli} for a similar derivation, without considering the slip
velocity):
\begin{eqnarray}
&&(\rho_g(L-z)+\rho_l z)
\ddot z  +  (\rho_l-\rho_g) (\dot z)^2  =  \nonumber \\
 &&2
  \frac{\gamma cos(\theta)}{H} - \frac{12 \dot z}{H^2(1+6 \frac{\lambda_s}{H})
} [(\mu_g (L-z) + \mu_l z)]\label{eq:front2}
\end{eqnarray}
In the above equation for the front dynamics, the terms in the LHS
take into account the fluid inertia. Being proportional either to the
acceleration or to the squared velocity, they become negligible for
long times. Washburn law plus the slip correction (\ref{eq:washburn1}) is therefore correctly recovered 
asymptotically, for $t \rightarrow \infty$, and in the limit when
$\rho_g/\rho_l \rightarrow 0$. The above equation is exact, in the
case where evaporation-condensation effects are negligible,
i.e. when the gas is pushed out of the capillary without any
interaction with the liquid. This is not always the case for most of
the mesoscopic models available in the literature \cite{shanchen,yeomans}, based on
a diffusive interface dynamics \cite{jacqmin}. As we will see, only
when either the limit of thin interface $\delta \xi / H \rightarrow 0$
is reached or when the gas phase is negligible, $\delta
\rho / \rho \rightarrow \infty$, the dynamics given by
(\ref{eq:front2}) is correctly recovered. Otherwise, deviations are observed, which
are induced by condensation/evaporation effects, which may result in significant
departure from the Poiseuille profile inside the gas phase.
\begin{figure}
  \resizebox{0.45\textwidth}{!}{
    \includegraphics{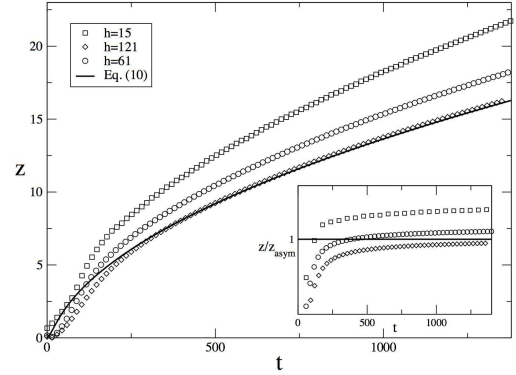}
  }
  \caption{Evolution of the front in adimensional variables. Position
    of the front, $z(t)$, versus $t$ for $H/(\Delta x)=15,61,121$  and
    $L/(\Delta x)=1200,1500,2000$ respectively, and with $\tilde
    z(0)=10 \Delta x$.  The solid
    curve is the theoretical solution obtained by integration of
    eq.(\ref{eq:front2}) with $\lambda_s=2$. The data have been
    obtained with ${\cal G}_b = 5$ which corresponds to $\rho_g=0.157$
    and $\rho_l=1.92$ (in LBE units). Let us notice that the solutions of
    eq.\ref{eq:front2} does not show any sensitive variations on $H$
    for those times and distances here explored. In the inset we show
    the position of the front for $(H/\Delta x)=15,61,121$ (same symbols)
    normalized with the adimensional asymptotic solution of
    eq.\ref{eq:front2} given by expression (\ref{eq:washburn1}):
    $z_{asym}(t) \sim \sqrt{\frac{cos(\theta)}{3}(1 + 6
      \frac{\lambda_s}{H}) t }$, notice that the departure from the
    predicted asymptotic Washburn law is never larger then $\%10$ even
    for small channel width, and becomes almost negligible already for
    $H=121 \Delta x$ }
  \label{fig:2}       
\end{figure}
\subsection{Numerical Results}
In fig.~(\ref{fig:2}) we show the behaviour of the front position,
$z(t)$, as a function of time for a given contact angle ($\theta=55^o \pm 3^o$),
a given density ration $\delta \rho/\rho = 11 $ and a given surface
tension $\gamma = 0.0569$ (in LBE units), at varying the channel width
$H$, from $H=15 \Delta x$ up
to $H=121 \Delta x$. As one can see the numerical results tends to be in good
agreement with the solutions of (\ref{eq:front2}), only for large
enough values of $H$, i.e. only when the interface becomes thin
enough.  For small to moderate values of $H$, the overall asymptotic
trend is only qualitatively reproduced by the Washburn law (plus slip
effects) (\ref{eq:washburn1}) with deviations which may be of the
order of $10\%$ in the prefactor (see inset of the same figure). 

\begin{figure}
  \resizebox{0.45\textwidth}{!}{
    \includegraphics{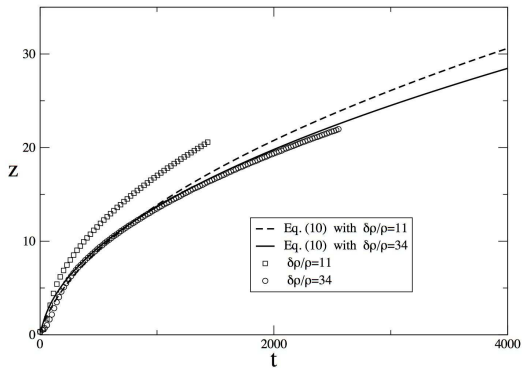}
  }
  \caption{Same plot of fig.~(\ref{fig:2})  but for fixed $H=31 \Delta
    x$ at changing the
    density ration. We have for ${\cal G}_b=5$, $\delta \rho/\rho = 11$
and for ${\cal G}_b=6$, $\delta \rho/\rho = 34$. Notice that the second case  is
    already enough to have a very good agreement with the solution of
    (\ref{eq:front2}) (adimensionalized as explained in the text) also at this small channel width (solid line ${\cal G}_b=6$, dashed line ${\cal G}_b=5$). }
  \label{fig:3}       
\end{figure}

Similarly, increasing the surface tension and the $\delta \rho/\rho$ factor, leads
to a early convergence towards the asymptotic Washburn law and to the
solution of (\ref{eq:front2}) even for small channel width $H$. This
is shown in  fig. (\ref{fig:3}) where, at fixed $H=31 \Delta x$, we
increase $\delta \rho/\rho$ and the Washburn law is approached 
better and better.

From both figure (\ref{fig:2}-\ref{fig:3}) one may notice that for
short filling time, $ t < \tau_{diff} $, strong deviations from the
Washburn law are detected, even considering the extra effects induced
by the inertial terms of (\ref{eq:front2}). This slowing down at
the early stage of the filling is indeed mainly induced by a sort of  {\it
vena contracta} term \cite{venacontracta}, reflecting the non-trivial matching between
the reservoir and the capillary dynamics at the inlet. This term, has
been argued to be describable by an additive  apparent-mass correction, 
$ c \rho_l  H \ddot z$,
to the LHS of   (\ref{eq:front2}). 

\begin{figure}
  \resizebox{0.45\textwidth}{!}{
    \includegraphics{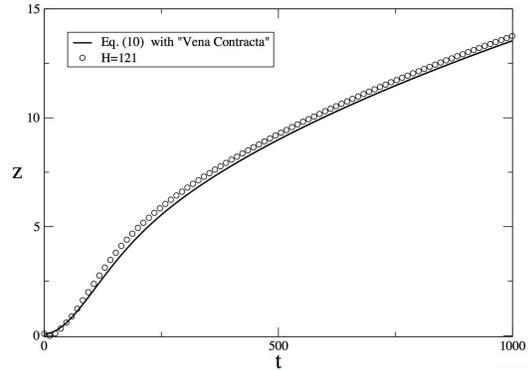}
  }
  \caption{Enlargement of  the early stage evolution during the filling process
for $H=121 \Delta x$ (in LBE units), with the vena contracta term choosing an optimal 
value for $c = 30$. The front position and time are make dimensionless 
normalizing with $t_{cap}$ and with $H$. }
\label{fig:4} 
\end{figure}

In Fig.~(\ref{fig:4}) we show an enlargement of
fig(\ref{fig:2}) for small filling time superposed with the results of
a numerical integration of eq. (\ref{eq:front2}) with a
phenomenological value $c=30$ for the vena contracta factor. As one
can see, the agreement between the numerics and the evolution of
(\ref{eq:front2}) is now excellent also at short times.
%
\section{Conclusions}
The present study shows that Lattice Boltzmann models with
pseudo-potential energy interactions are capable of reproducing the
basic features of capillary filling, as described within the Washburn
approximation.  Two conditions for quantitative agreement have been
identified: i) a sufficiently high density contrast between the
dense/light phase, $\rho_l/\rho_g > 10$ and a sufficiently thin
interface, $\delta \xi/H < 0.1$.  Both conditions can be met within
the current LB methodology, although it would clearly be desirable to
extend the LB scheme in such a way to achieve density contrasts in the
order of $1:1000$ (the current state-of-the-art is approximately
$1:50$) and interface widths of the order of the lattice spacing
$\Delta x$ (current values are about $5-10 \Delta x$).

The present results set the stage for future computational studies
aimed at identifying optimal interface functionalization strategies,
based on physical, chemical and geometrical coating processes.  Work
along these lines is currently underway.
\section{Acknowledgments}
Valuable discussions with E. Paganini, D. Palmieri and D. Pisignano
are kindly acknowledged. Work performed under the EC contract NMP3-CT-2006-031980 (INFLUS).

\end{document}